\documentclass[12]{article}
\usepackage{cite}
\usepackage{lineno}
\usepackage{amsmath}
\usepackage{amsfonts}
\usepackage{amssymb}
\usepackage{graphicx}
\usepackage{subcaption}
\usepackage{ragged2e}
\usepackage{hyperref}
\DeclareCaptionJustification{justified}{\justifying}
\makeatletter
\renewcommand{\maketitle}{\bgroup\setlength{\parindent}{0pt}
\begin{flushleft}
  {\Large\bfseries\noindent\sloppy \textsf{\@title} \par}

  \@author
\end{flushleft}\egroup
}
\makeatother

\newenvironment{affiliations}{%
    \setcounter{enumi}{1}%
    \setlength{\parindent}{0in}%
    \slshape\sloppy%
    \begin{list}{\upshape$^{\arabic{enumi}}$}{%
        \usecounter{enumi}%
        \setlength{\leftmargin}{0in}%
        \setlength{\topsep}{0in}%
        \setlength{\labelsep}{0in}%
        \setlength{\labelwidth}{0in}%
        \setlength{\listparindent}{0in}%
        \setlength{\itemsep}{0ex}%
        \setlength{\parsep}{0in}%
        }
    }{\end{list}\par\vspace{12pt}}

\renewcommand{\mathbf}{\boldsymbol}

\usepackage{ulem}
\title{Coherence resonance in influencer networks}
\author
{Ralf T\"onjes$^{1}$,   Carlos E. Fiore$^{2}$ and Tiago Pereira$^{3,4}$}

\date{}

\begin{document}

\maketitle

\begin{affiliations}
\item Institute for Physics and Astronomy, University of Potsdam, Karl-Liebknecht-Str. 24, 14476 Potsdam, Germany
\item Instituto de F\'isica, Universidade de S\~ao Paulo, S\~ao Paulo, Brazil
\item Instituto de Ci\^encias Matem\'aticas e Computa\c{c}\~ao, Universidade de S\~ao Paulo, S\~ao Carlos, S\~ao Paulo, Brazil
\item Department of Mathematics, Imperial College London, SW7 2AZ, London, UK
\end{affiliations}

\section*{Abstract}
{\bf 
Complex networks are abundant in nature and many share an important structural property: they contain a few nodes that are abnormally highly connected (hubs). Some of these hubs are called influencers because they couple strongly to the network and play fundamental dynamical and structural roles. Strikingly, despite the abundance of networks with influencers, little is known about their response to stochastic forcing. 
Here, for oscillatory dynamics on influencer networks, we show that
 subjecting influencers to an optimal intensity of noise can result in enhanced network synchronization.  This new network dynamical effect, which we call coherence resonance in influencer networks, emerges from a synergy between network structure and stochasticity and is highly nonlinear, vanishing when the noise is  too weak or too strong. Our results reveal that the influencer backbone can sharply increase the dynamical response in complex systems of coupled oscillators.
}

\section*{Introduction}

A central discovery in network science is that a small group of highly connected hubs can couple to the network more strongly than their peers and greatly influence the network behavior \cite{Newman,xu,WSNature,Heuvel,Song,fulcher}. Examples of network influencers can be found in neuroscience (e.g., normal and aberrant synaptic connectivity \cite{HubScience,Nat2,buzsaki2014log,Gal}), political opinions (e.g., election blogging \cite{Adamic} or social networks), and man-made scale-free networks (e.g., the internet \cite{Newman}). Surprisingly, the presence of such influencers makes synchronization of deterministic network dynamics more difficult because networks with influencers require stronger coupling than homogenous networks \cite{TiagoPRL,Thomas}; indeed, in many situations, synchronization of influencer networks cannot be achieved at all \cite{JEMS,Adilson}. This observation is all the more remarkable because synchronization plays a fundamental role in regulating network function \cite{Fries,Bruno} and is mediated predominantly through influencers \cite{HubScience,Morgan,zamora}. This raises a crucial question: why have many real-world networks evolved to contain influencers when they appear to be detrimental to the network dynamics, at least at face value?

While strong random fluctuations usually have a negative effect in complex systems it has long been recognized that a small amount of noise can actually improve the system response and its ability to process information. Known mechanisms for such a constructive influence of noise are stochastic resonance, coherence resonance and noise induced synchronization \cite{Lindner,Luca,Arkady1,wang2000,Semenova,masoliver2017,ushakov2005,Lindner2,Nakao,Pimenova,GoldobinE}. The term coherence resonance is used to describe an optimal response of noise-induced oscillations  without external stimmulus in excitable cells \cite{Arkady1}. It was  observed in globally coupled systems \cite{wang2000}, in homogeneous networks \cite{Semenova,masoliver2017},  in non-excitable systems near a Hopf bifurcation \cite{ushakov2005} and two coupled oscillators \cite{Lindner2}. The effects of coherence resonance and its role in heterogeneous networks such as influencer networks remains elusive.

In this work, we show that stochastic forcing of influencers can lead to an optimal collective network response. Strikingly, introduction of noise synergizes with the network structure to create collective oscillations that become optimal at a given noise strength in the influencers. This phenomenon emerges in two steps. First, the network acts as a nonlinear filter for the stochastic influencer dynamics, and at an optimal noise strength, the influencers induce synchronization in the nodes directly connected to them. Second, different parts of the network develop macroscopic dynamics and interact indirectly through the influencers. We develop an adiabatic theory to uncover this macroscopic interaction law and show that it mediates the emergence of global collective oscillations. When the noise in the influencers is either too weak or too strong, the coupling vanishes. Interestingly, at a macroscopic level, the interaction between different parts of the network can be described by a hyper-graph. 

We refer to a network where most nodes couple predominantly to a small number of influencers as an influencer network, and refer to the remaining nodes as followers (Figure 1). As generic oscillatory dynamics, we consider a network of phase oscillators
\begin{linenomath*}
\begin{equation}\label{standard}
\dot\vartheta_n = \omega_n + \frac{\lambda_n}{\mu_n} 
\sum_{m=1}^{N} W_{nm} g(\vartheta_m,\vartheta_n) 
+\sqrt{2D_n} \xi_n.
\end{equation}
\end{linenomath*}
Here, $\omega_n$ is the natural frequency of node $n$, which couples with strength $\lambda_n$ to the weighted mean of the coupling functions $g$ to neighboring nodes $m$. Given a network weight coupling matrix $W_{nm}\ge 0$, which is nonzero if node $n$ receives a link from node $m$, the intensity $\mu_n$ is the total coupling weight received by the $n$th node. A table of parameters and their function is provided in Methods. The coupling 
\begin{linenomath*}
\begin{equation}
g(\vartheta_m, \vartheta_n) = \sin \left( \vartheta_m - \vartheta_n - \alpha \right) 
 + c_0
\end{equation}
\end{linenomath*}
is generic for weakly coupled, nearly identical oscillators \cite{KuraModel,RMP}. The parameter $\alpha$ is called phase frustration and the bias $c_0$ is due to shear, an amplitude dependence of the frequency \cite{Montbrio,Bard}.  The effect of shear is a shift in the average frequency proportional to the coupling strength. Phase equations with this form of coupling $g$ are known as the Kuramoto-Sakaguchi model \cite{sakaguchi1986,omel2012} and are widely applied across scientific disciplines \cite{KuraModel,RMP,Montbrio,Bard,sakaguchi1986,omel2012}.  In addition, each term {$\sqrt{2D_n}\xi_n$} denotes uncorrelated Gaussian white noise of strengths $D_n$. In many studies, the coupling strength $\lambda_n$ to the local mean-field is chosen to be uniform, in which case the coupling is called normalized. In real-world and experimental systems, though, coupling may be heterogenous and hubs can couple more strongly to the network \cite{Song,Morgan}. We model this coupling as 
\begin{linenomath*}
\begin{equation}
\lambda_n = 
\left\{
\begin{array}{cl}
\beta_n \lambda_0 & \mbox{~for~influencers} \\
\lambda_0  & \mbox{~for~followers.}
\end{array}
\right. 
\end{equation}
\end{linenomath*}
For simplicity, throughout this exposition we consider the coupling intensity $\beta_n=\beta$, the noise strength $D_n=D$, and the frequency $\omega_n=\omega$, to be identical for all influencers. For the followers we assume a Lorentzian frequency distribution with mean frequency $\omega_0$ and width $\gamma_0$. Noise  is of identical strength $D_n=D_0$ in all followers.  We denote $\Delta\omega=\omega-\omega_0$ the average gap in natural frequencies between influencers and followers.

In Methods we show how  Eq. \eqref{standard} can be recast in terms of  dimensionless effective parameters shown in Table \ref{tab1}. These effective parameters, and in particular the influencer effective noise strength $q=D/\Delta\Omega$, play key roles in the collective dynamics of the system. The dynamical frequency gap $\Delta\Omega/\lambda_0$ leads to a time scale separation between the dynamics of the followers and  the influencers. A coupling intensity $\beta$ of comparable but smaller magnitude leads to an effective coupling strength $\Lambda$ close to one for which the effect of coherence resonance is most pronounced. We note that the dynamical frequency gap needs to be large in units of $\lambda_0$, but it can be small in natural time units. In Supplementary Note 1, we present an example of the transformation for realistic parameters in Eq. (\ref{standard}) to effective parameters.

\begin{table}[t!]
\centering
\begin{tabular}[t]{lll}
\hline
Parameter  & Meaning & range\\
\hline
\hline
$\Delta\Omega/\lambda_0 = \Delta\omega/\lambda_0 + (\beta-1)c_0$ 
&  dynamical frequency gap &  {$\Delta\Omega/\lambda_0\gg 1$} \\
$ \Lambda = \beta\lambda_0/\Delta\Omega$ & dimensionless coupling strength &  {$\Lambda < 1$} \\
$q = D/\Delta \Omega$ & influencer effective noise strength &  {$q =  O(1)$} \\
$D_0/\lambda_0$  & followers effective noise strength & {$D_0/\lambda_0 \ll 1$}\\
$\gamma_0/\lambda_0$ & followers frequency heterogeneity & {$\gamma_0/\lambda_0\ll 1$}\\
\hline
\end{tabular}
\caption{{\bf Effective dynamical parameters in influencer networks}.  
We obtain these parameters as described in Methods. 
These are key parameters in the description of 
coherence resonance and the optimal noise strength in the influencers.
}
\label{tab1}
\end{table}

We divide the followers into partitions $P_{\sigma}$ of nodes connected to the same set of influencers. In Figure 1, we show an influencer network with two influencers ($a$ and $b$) and three partitions of followers ($\sigma$=1,2,3), which are connected to influencers $a$, $b$, or both (see additional examples in Supplementary Note 2). To capture the collective dynamics in each partition $\sigma$, we introduce the complex mean-fields
\begin{linenomath*}
\begin{equation}\label{OrderPar}
Z_{\sigma} (t)= \frac{1}{|P_{\sigma}|} \sum_{n \in P_{\sigma}} 
e^{i \vartheta_n (t) }
\end{equation} 
\end{linenomath*}
The modulo of the complex mean-field 
$R_{\sigma} = |Z_{\sigma}|$ is the partition order parameter; that is, $R_{\sigma} = 0$ for incoherent, uniformly distributed phases and $R_{\sigma} = 1$ in full synchrony. Similarly, the global mean-field $Z$ and order parameter $R$ are defined by summing over all followers in the network.

\begin{figure}[t!]
\centering
\includegraphics[width=12cm]{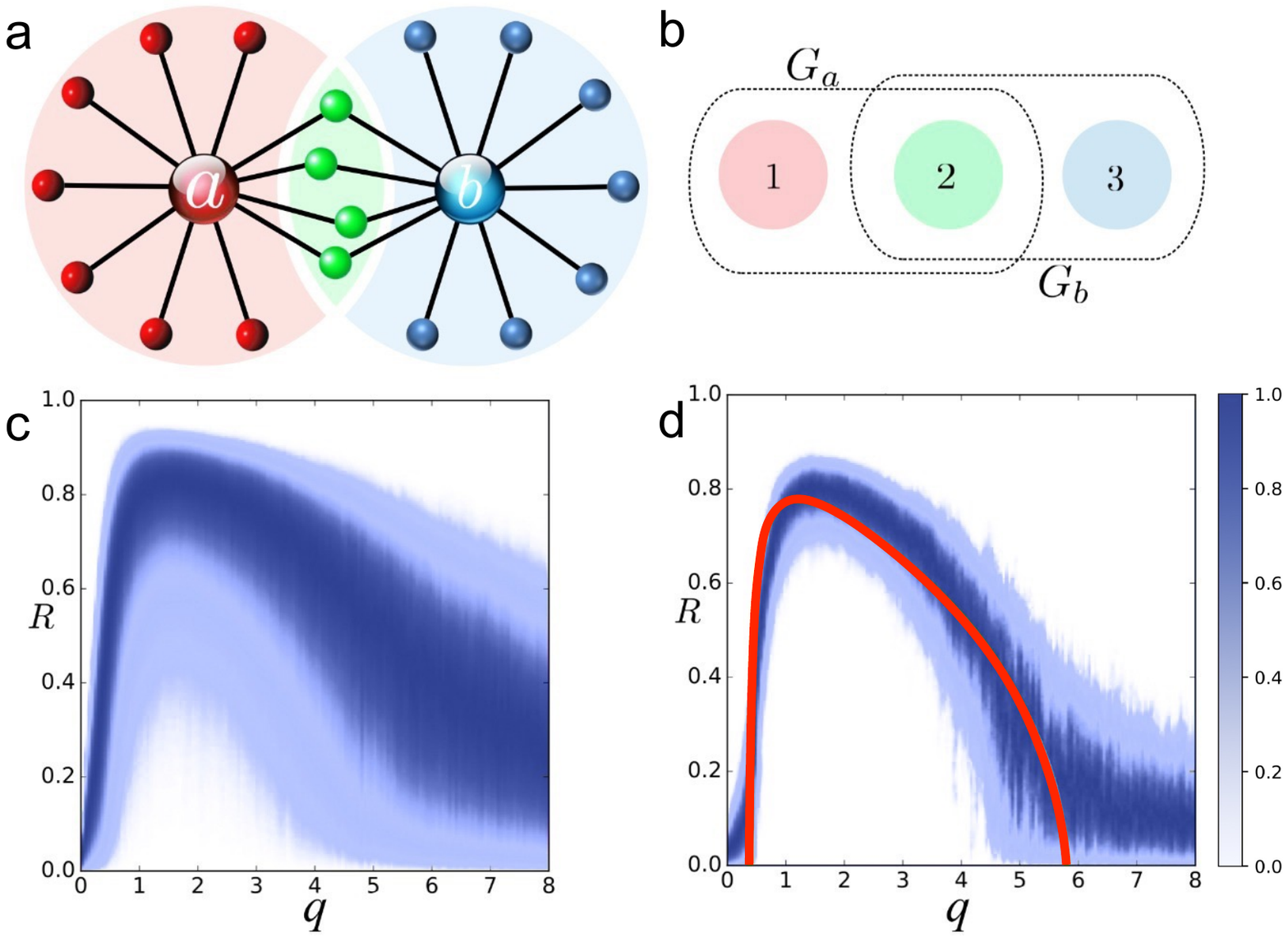}
\caption{
{\bf Coherence resonance in an influencer network.
} Distribution of the order parameter $R$ versus 
the effective diffusion $q$ in the influencers. (a)
Influencers $a$ and $b$ are hubs that couple strongly to the network, and all other nodes are regarded as followers. Three distinct partitions of followers are shown in red, blue, and green, which connect to influencer $a$, $b$, and both $a$ and $b$, respectively. In our simulation, each partition has 300 followers. (b) Mean-field theory predicts that the interactions of the partition mean-fields take place in a hyper-graph mediated by the coupling functions $G_a$ and $G_b$  see Methods. (c,d) For each value of effective noise strength $q$ in the influencers, we plot the density of the global order parameter $R$ on a color scale from $0$ (white) to the maximum value (dark blue). At an optimal noise strength, the mean value of the global order parameter reaches a maximum, revealing the coherence resonance effect. In (c) the dynamical frequency gap between influencers and followers $\Delta\Omega/\lambda_0=18$ is moderate, whereas in (d) $\Delta\Omega/\lambda_0=198$ is large. The solid red line in (c) is our analytical prediction.}
\label{Fig1}
\end{figure}

\section*{Results}  

With deterministic influencers where $q=0$, and when $|\Lambda|<1$, the influencers cannot frequency lock to the followers. Synchronization of the followers through the influencer backbone is poor and counteracted by noise and frequency heterogeneity in the followers. Our results show that by setting a weak noise strength or frequency heterogeneity in the followers and by changing the effective noise $q$ in the influencers, synchronization of the whole network increases, reaches a maximum, and then decreases.

We numerically integrate our model Eq. (\ref{standard}) in dimensionless units (Table \ref{tab1}) for the network with two influencers (as shown in Figure 1) with $300$ identical followers in each partition, and a small fixed noise strength in the followers. {By changing the noise strength in the influencers}, we then obtain the distribution of the order parameter $R$ as a function of $q$. {After a transient, the order parameter is independent of the initial conditions.} At an optimal noise strength, $R$ reaches its maximum (in expected value), as shown in Figure 1 for $\Delta\Omega/\lambda_0 = 18$ (panel c) and for $\Delta\Omega/\lambda_0 = 198$ (panel d). The solid line is a theoretical prediction in the thermodynamic limit for heterogeneous followers using a slow-fast approximation. Frequency heterogeneity and noise in the followers have qualitatively and quantitatively the same desynchronizing effect. Optimal synchronization of the whole network is predicted theoretically and achieved in all simulations for an effective noise strength $q\approx1$ in the influencers, see details in Methods.  Our mean-field analysis predicts that the effect of coherence resonance is only observed for very small frequency heterogeneity or noise in the followers, below a threshold that depends on $\Lambda$ (Supplementary Note 3).

In Figure 2, we show the time series of the order parameter $R$ for two complex and real-world networks. The upper row represents a scale-free network and the lower row the directed neural network in the model organism {\em Caenorhabditis elegans}. We assign the role of influencers to the $K$ most strongly connected nodes and use a weighted connectivity matrix $W_{nm}=1$ for all connections from or to an influencer, and $W_{nm}=0.01$ for all other connections. For small effective noise $q$ in the influencers (Figure 2, $q$ weak), the order parameter fluctuates at a low level. When $q =1$ ($q$ optimal), the order parameter fluctuates around a value close to $1$, revealing coherent collective oscillations. Finally, when $q$ is large ($q$ strong), the order parameter decreases again, revealing the loss of synchrony. All parameters for the simulation and numerical scheme can be found in Methods. In Supplementary Note 4, we show three additional examples of coherence resonance in influencer networks;  with 3 influencers, a random network with $100$ influences, and a network of linked political blogs. 

\begin{figure}[t!]
\centering
\includegraphics[width=12cm]{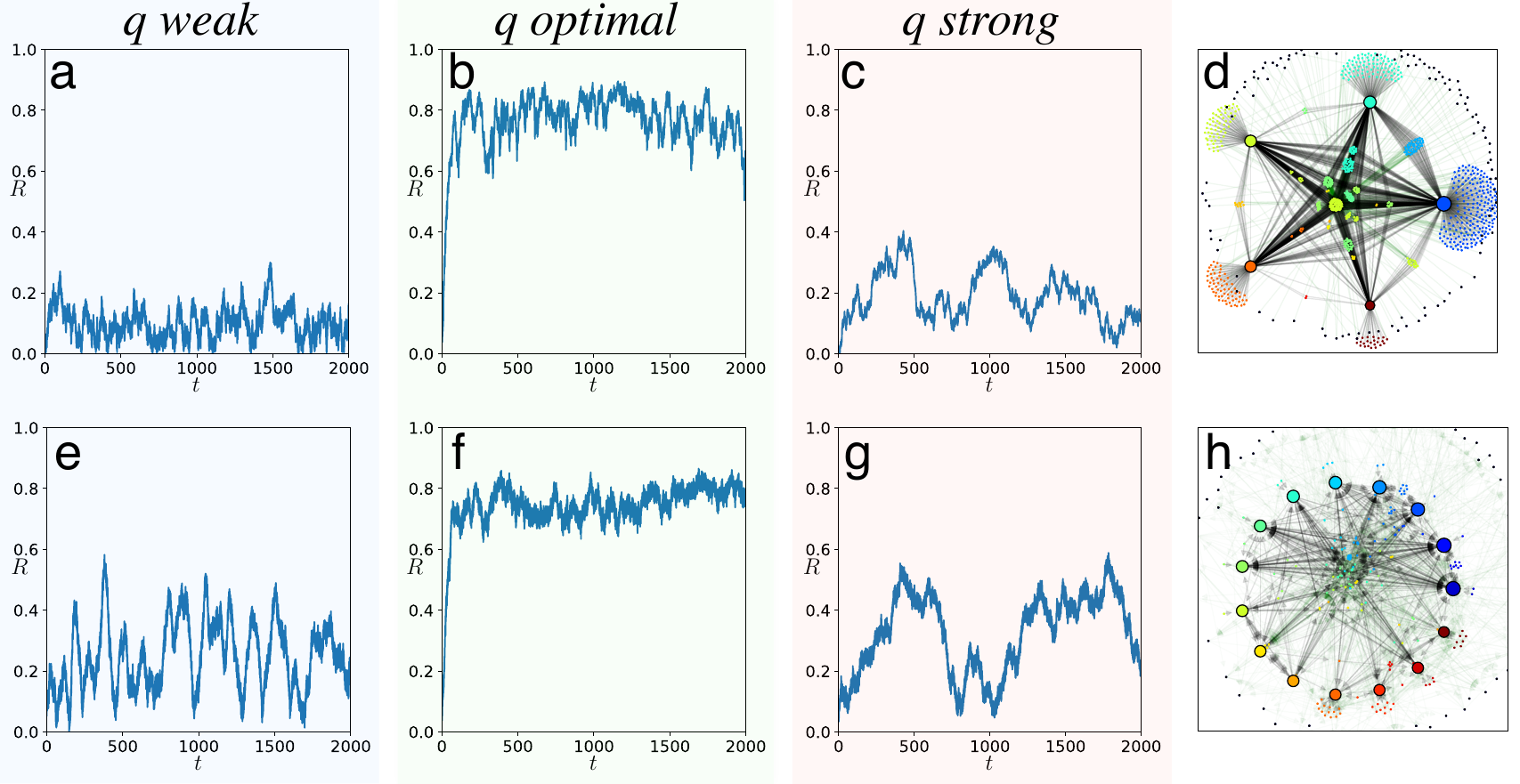}
\caption{
{\bf Coherence resonance of the order parameter in different complex networks}. (a-c, e-g) show the time series of the order parameter $R$ for three values of noise strength in the influencers for weak $q =0.1$ (a,e), optimal $q=1$ (b,f), and strong q=10 (c,g). (d,h) show the corresponding networks with (d) a scale-free network with exponent $2$ \cite{Newman} and (h) {\em C. elegans} directed neural network \cite{WSNature}.  See Methods for further details. Additional examples can be found in Supplementary Note 5.
}
\label{Fig2}
\end{figure}

\subsection*{Stochastic forcing by a single influencer} 

Let us consider a single influencer. When its followers are asynchronous, the sinusoidal contributions in the sum of the coupling functions for that influencer average out, and the influencer phase is effectively decoupled from the followers. The influencer is independent of the network and acts as a common stochastic force on the followers connected to it. The additive noise in the influencer enters the dynamics of its followers multiplicatively through the coupling function. That is, the network acts as a nonlinear filter for the noise in the influencers. We can show that the effective diffusion constant of the integrated stochastic forcing in the followers, a proxy for the noise strength, attains a maximum at an optimal effective noise strength
\begin{linenomath*}
\begin{equation}\label{qopt}
q_{\rm opt }=1\,.
\end{equation}
\end{linenomath*}
We present the calculations in Methods and further details in Supplementary Note 5.
When oscillations are driven by common multiplicative noise, the effect of noise-induced synchronization can be observed \cite{Nakao,Pimenova,GoldobinE}. As the common noise intensity is increased the oscillators synchronize faster. This suggests that at this optimal value of $q$ the incoherent state will be most unstable. However, behavior of the noise transfer does not explain the synchronization between different partition mean-fields. Because this synchronization requires studying macroscopic dynamics of $R$ far from zero, our next goal is to uncover the interaction function between the mean-fields in different partitions.

\subsection*{Mean-field dynamics of partitions takes place in a hyper-network} 

For simplicity, we provide an analysis of the influencer network shown in Figure \ref{Fig1}. Mean-field equations for mixed repulsive and attractive coupling or intra- and inter-partition interactions in the followers can be generalized from these results. In Methods, we show that assuming large partition sizes  $|P_{\sigma}| \gg 1$, large dynamical frequency gap $\Delta\Omega/\lambda_0 \gg 1$, and noise free followers $D_0=0$ with frequency heterogeneity $\gamma_0$, it is possible to derive averaged dynamics of the partition mean-fields $Z_\sigma$ in an adiabatic approximation. The resulting deterministic equations have the structure of a hyper-network. In our example, the governing equations are
\begin{linenomath*}
\begin{equation}\label{Eq:Example_MFeq}
\dot{Z}_1 = F (Z_1, G_a),  \quad \dot{Z}_2 = 
F \left(Z_2, \frac  1 2 G_a + \frac 1 2  G_b  \right)\quad 
\mbox{and} \quad \dot{Z}_3 = F (Z_3, G_b) 
\end{equation}
\end{linenomath*}
where the coupling functions are
\begin{linenomath*}
\begin{equation}\label{Eq:Example_G}
G_a = G(w_{a1} Z_1 + w_{a2} Z_2) \mbox{~~ and ~~} 
G_b = G(w_{b2} Z_2 + w_{b3} Z_3).
\end{equation}
\end{linenomath*}
$F$ describes a Riccati force (see Methods). The weights 
$w_{k\sigma}$ with $\sum_\sigma w_{k\sigma}=1$ denote the relative size of partition $\sigma$ among all followers of influencer $k$. Note that, while in the microscopic description the connections between nodes are pairwise, at the level of mean-fields, edges represented by a coupling function $G_k$ for each influencer can connect multiple partitions of followers. Thus, the mean-field interaction between different parts of the network is described by a hyper-graph.

The interaction functions $G_a$ and $G_b$ can be determined analytically; they depend on $\Lambda$ and $q$, are maximal at an optimal noise strength, and vanish at critical values of $q$. That is, at weak or strong  noise in the influencers, the hyper-network interactions vanish, revealing the highly nonlinear nature of the phenomenon. In particular, this means that the macroscopic fields will not interact in the strong noise limit. We derive the analytic expressions for the coupling functions $G_k(Z;\Lambda,q)$ in Methods.

\subsection*{Global synchronization and resonance} 
When influencers have equal parameters $q$ and $\Lambda$ the synchronization manifold $Z_\sigma=Z$ is invariant under (\ref{Eq:Example_MFeq}) and (\ref{Eq:Example_G}) and stable for phase-attractive coupling. Hence, the macroscopic fields synchronize and we can explain the global coherence resonance by restricting the analysis to this invariant subspace. Our mean-field theory predicts both effects: the coherence resonance of the partition order parameters and phase synchronization of the partition mean-fields as shown in Methods. The solid line in Figure \ref{Fig1} (right) is the stationary average order parameter predicted by our theory in the infinite time-scale separation limit and with frequency heterogeneity $\gamma_0/\lambda_0=0.02$  in the followers.  The predicted values agree with simulations of the finite size network, large dynamical frequency gap $\Delta\Omega/\lambda_0=198$ and identical followers with noise $D_0/\lambda_0=0.02$. In Supplementary Information, we provide two short movies displaying synchronization of the network in Fig.1 at an optimal noise strength in the two influencers.

\section*{Discussion} 
We have found a new effect induced by a synergy between noise and network structure to generate a transition towards a synchronization that would not be possible in the absence of noise. The key element for this effect is the existence of influencers -- a group of hubs that couple strongly and connect different parts of a network. Although deterministic network parameters prevent synchronization, we show that an optimal noise strength in the influencers can induce and mediate synchronization. The mechanism for this coherence resonance in influencer networks is different from the known effect of coherence resonance in homogeneous networks with excitatory dynamics, where noise simply excites oscillations \cite{wang2000,Semenova,masoliver2017}. At the macroscopic level, the interaction between different parts of the network is indirect and takes place on an emerging hyper-network, thus changing the interaction structure from the microscopic level. Such higher order interactions have previously been conjectured and reported in neuronal data recordings \cite{Basset16}. Our findings suggest that the emergent order in complex systems could be controlled by regulating the noise in only a few key nodes.
\section*{Methods}
\subsection*{Canonical form} 
To bring the  Eq. (\ref{standard}) into a dimensionless form with effective parameters given in Table \ref{tab1},  we change the time scale to units of $1/\lambda_0$ and add the frequency shift from the bias $c_0$ in the coupling function to the natural frequencies of the oscillators, i.e. $\omega_n \mapsto \omega_n + \lambda_n c_0$ and $g(\vartheta_m,\vartheta_n)\to \sin(\vartheta_m-\vartheta_n-\alpha)$. The difference between the average follower frequency and the frequency of an influencer, both including the frequency shift from the coupling bias, is the dynamical frequency gap $\Delta\Omega/\lambda_0$ (in units of $\lambda_0$). Observing the invariance of the phase equations under a global phase shift, i.e. $\vartheta\to\vartheta-\omega_0 t$, we can go into a co-rotating reference frame where the average follower frequency is zero. The deviations of the follower frequencies from their mean frequency $\omega_0$ may be written as $\gamma_0\nu_n$ where the $\nu_n$ are taken from some standard distribution with mean zero and the factor $\gamma_0\ge 0$ characterizes the frequency heterogeneity.
Then the phase equations for the followers in the new time units and co-rotating reference frame are
\begin{linenomath*}
\begin{equation}\label{Eq:CanonicFollowers}
	\dot\vartheta_n =\frac{\gamma_0}{\lambda_0} \nu_n + \frac{\lambda_n}{\lambda_0}\frac{1}{\mu_n}\sum_k W_{nk}\sin(\vartheta_k-\vartheta_n-\alpha) + \sqrt{2D_n/\lambda_0}\tilde{\xi}_n
\end{equation}
\end{linenomath*}
and for the influencers with phases $\psi_k$
\begin{linenomath*}
\begin{equation}\label{Eq:CanonicInfluencers}
	\dot\psi_k = \frac{\Delta\Omega_k}{\lambda_0}\left(1 + \Lambda_k \frac{1}{\mu_k}\sum_m W_{km} \sin(\vartheta_m-\psi_k-\alpha) \right) + \sqrt{2q_k\frac{\Delta\Omega_k}{\lambda_0}}\tilde{\xi}_k.
\end{equation}
\end{linenomath*}
Here, $\Lambda_k = \lambda_k/\Delta\Omega_k$ is the ratio between the coupling strength and the frequency of the influencer. In the noise free case, phase locking is only possible for $|\Lambda_k|>1$.  Changing $\Lambda$ can lead to a discontinuous, explosive synchronization \cite{TiagoPRL,Vova}.  The terms $\tilde{\xi}_m$ are independent white noise with $\langle\tilde{\xi}_m(t)\tilde{\xi}_n(t')\rangle = \delta_{mn}\delta(t-t')$ in the new units of time and $q_k$ is the effective noise strength in the influencers on the fast time scale $\Delta\Omega_k/\lambda_0$. In Supplementary Note 1, we provide examples of such rescaling. 

\subsection*{Parameters and their meaning}
In Table \ref{tab2}, we present the main parameters that naturally appear in the phase model Eq. (1) and give rise to effective parameters, as shown in Table 1 in the main text. The main parameters in our mean-field analysis are shown in Table \ref{tab3}.
\begin{table}[!h]
\centering
\begin{tabular}[t]{ll}
\hline
Parameter  & Meaning \\
\hline
\hline
$\omega_n$ &  isolated frequency of the $n$th oscillator; \\
	& set as $\omega_n=\omega_0+\gamma_0\nu_n$ for followers and $\omega$ for influencers\\
$\omega_0$ & mean frequency of the followers \\
$\gamma_0\nu_n$ &  frequency deviation $\omega_n-\omega_0$ of the $n$th follower\\
$\gamma_0$ &  scale parameter of follower frequency distribution  \\
$\Delta \omega$ &   gap $(\omega-\omega_0)$ between influencer and average follower frequency\\
$W_{nm}$ & nonnegative matrix of connection weights \\
$\mu_n$ & connection intensity ( $\mu_n = \sum_m W_{nm}$) \\
$\lambda_n$ &  coupling strength of the $n$th oscillator;\\
	& $\lambda_0$ for followers and $\beta \lambda_0$ for influencers\\
$\beta$ & coupling intensity for influencers\\ 
$D_n$ &  noise strength set as $D$ for influencers and $D_0$ for followers \\
$\alpha$ &    phase frustration in the coupling function $g$ \\
$ c_0$ &  shear parameter in the coupling function $g$ \\
\hline
\end{tabular}
\caption{{\bf Parameters in the model presented in Eq. (1) of the main manuscript}.}
\label{tab2}
\end{table}
\begin{table}[!h]
\centering
\begin{tabular}[t]{ll}
\hline
Parameter & Meaning \\
\hline
\hline
$P_\sigma$ &  follower partitions according to the influencers they connect to \\
$I_\sigma$ &  set of influencers of a partition $\sigma$\\
$Z_{\sigma}$ &  complex mean-field of partition $\sigma$ (order parameter $R_\sigma = |Z_\sigma|$)\\
$G_k$ &  coupling function between mean-fields mediated by influencer $k$\\
$w_{k \sigma}$ &   relative size of partition $\sigma$ among the followers of influencer $k$ \\
$F$ & Ricatti vector field  see Eq. (\ref{Eq:OA_CommunityDyn}) \\
$h_\sigma$, $h_k$  &   forces on oscillators in partition $\sigma$ and on influencer $k$\\
$H_\sigma$ &  average of $h_\sigma$ obtained from adiabatic mean-field approximation\\
\hline
\end{tabular}
\caption{{\bf Parameters of the mean-field analysis presented in Eq. (\ref{Eq:Example_MFeq}).}}
\label{tab3}
\end{table}
\subsection*{Simulations and parameter values}
In our analysis and our simulations, we use the transformed, dimensionless canonical form (\ref{Eq:CanonicFollowers}) and (\ref{Eq:CanonicInfluencers}) of the phase equations (\ref{standard}).  Existing connections in the network from and to the influencers are given the weight $W_{mn}=1$ and connections between followers $W_{nm}=0.01$. The followers couple to their neighbors with strength $\lambda_n=\lambda_0$ and influencers with strength $\lambda_k=\beta\lambda_0$. The phase frustration in the coupling function is set to $\alpha=-0.1$. The frequency deviations $\nu_n$ of the followers are drawn from a Cauchy distribution $p(\nu)=1/\pi(\nu^2+1)$ and multiplied by $\gamma_0/\lambda_0$. Thus, frequency heterogeneity and noise strength in the followers are given by $\gamma_0/\lambda_0$ and $D_0/\lambda_0$, respectively. We chose $\beta_k=\beta$,  $\lambda_k=\beta\lambda_0$ and $D_k=D$ for all influencers, so that $\Delta\Omega/\lambda_0$, $\Lambda$ and $q$ are identical for all influencers. We integrate the Langevin equations of the phases with an Euler-Maruyama scheme and small time steps $dt=5\cdot10^{-4}$ because of the large time scale separation $\Delta\Omega/\lambda_0\gg 1$. 

The parameters of Figure 1 are as follows: The network structure is a pure influencer network without connections between followers or between influencers. We simulate 300 identical followers $\gamma_0/\lambda_0=0$ in each of the three partitions with small independent noise $D_0/\lambda_0=0.02$. In the lower left panel we have $\beta=10$, a dynamical frequency gap of $\Delta\Omega/\lambda_0 = 18$ and an effective influencer coupling strength $\Lambda=10/18$. In the lower right panel $\beta=100$, $\Delta\Omega/\lambda_0 = 198$ and $\Lambda=100/198$. For each value of the effective influencer noise strength $q=D/\Delta\Omega$ we record a histogram of the order parameter over $T=10^4$ time units, {which is much longer than the relaxation time of $R$.} The theoretical prediction, the solid line in the lower right panel, is for noiseless followers $D_0=0$ and $\gamma_0/\lambda_0 = 0.02$. 

Parameters of Figure 2 are as follows: For the {\it C. elegans} directed neuronal network \cite{WSNature}, we choose the top $K=15$ out-degree nodes as influencers. All nodes with zero in-degree have been removed, resulting in a network with $N=268$ nodes. Connections between followers are given the weight $W_{nm}=0.01$. We simulate identical followers $\gamma_0/\lambda_0=0$ with small independent noise $D_0/\lambda_0=0.02$. The dynamical frequency gap between followers and influencers is $\Delta\Omega/\lambda_0=18$ and the effective coupling strength in the influencers is $\Lambda=10/18$. Shown are three time series of the network order parameter for small ($q=0.1$), optimal ($q=1$), and large ($q=10$) noise strength in the influencers. The undirected scale-free network with exponent 2 is the largest connected component of a network  generated via a configurational algorithm \cite{Newman} without self loops or double edges. We chose the top 5 degree nodes as influencers. The other parameters are the same as in the {\it C. elegans} neuronal network.
\subsection*{Mean-field dynamics in influencer networks}
We have developed a mean-field theory for undirected influencer networks with connections exclusively between influencers and followers, as shown in Figure \ref{Fig1}. This theory can be generalized to more complex configurations, heterogenous influencers, directed, attractive, or repulsive coupling between followers and influencers, within partitions or between different partitions. While these generalizations may lead to more complex dynamic behavior, the mechanism for the coherence resonance is apparent in the simplest model.

We consider the network as a union of a set $P$ of followers and a set $I$ of influencers.  The nodes $n$ connected to an influencer $k$ are elements $n\in P_k$ of the periphery of the influencer $k$. Intersections of the sets $P_k$ form equivalence classes or partitions $P_\sigma$ of followers that are connected to the same subsets $I_\sigma$ of influencers such as in Fig.\ref{Fig1} all followers connected to influencer $a$ or $b$ or to both influencers. {The phases of the oscillators are encoded as complex variables} $z_n = \exp(i\vartheta_n)$ for the followers  and $z_k = \exp(i\psi_k)$ for the influencers.  The dynamics can be formulated in terms of partition averages and averages over the influencers of these partitions
\begin{linenomath*}
\begin{eqnarray} 
Z_\sigma &=& \frac{1}{|P_\sigma| } \sum_{n\in P_\sigma} z_n \label{Eq:Fields01}\\
h_\sigma &=& \frac{e^{-i\alpha}}{2i} \frac{1}{|I_\sigma|}
\sum_{k\in I_\sigma} z_k \label{Eq:Fields02} \\
h_k&=& \frac{e^{-i\alpha}}{2i}  \frac{1}{|P_k|} \sum_{n\in P_k} z_n = 
\frac{e^{-i\alpha}}{2i} \sum_{\sigma} w_{k\sigma} Z_\sigma. \label{Eq:Fields03}
\end{eqnarray}
\end{linenomath*}
Here, $h_\sigma$ and $h_k$ are the forces acting on the followers in partition $\sigma$ and on the influencer $k$. The weight $w_{k\sigma}$ is the relative size of partition $\sigma$ in the periphery of an influencer $k$; that is,  $w_{k\sigma} = |P_\sigma| / |P_k|$ when $P_\sigma\subseteq P_k$ or $w_{k\sigma}=0$ otherwise. Thus, the phase dynamics (\ref{Eq:CanonicFollowers}) and (\ref{Eq:CanonicInfluencers}) can be written in complex form as
\begin{linenomath*}
\begin{eqnarray}
\dot z_n &=& iz_n \left(\bar{h}_\sigma z_n + \frac{\gamma_0}{\lambda_0} 
\nu_n + h_\sigma \bar{z}_n\right) + iz_n\sqrt{2D_0/\lambda_0}\xi_n(t), 
\qquad n\in P_\sigma  \\
\dot z_k &=& iz_k\frac{\Delta\Omega_k}{\lambda_0}
\left(\Lambda_k \bar{h}_k z_k + 1 + \Lambda_k h_k \bar{z}_k\right) 
+ iz_k\sqrt{2D_k/\lambda_0}\xi_k(t) \label{Eq:HubDynamics}
\end{eqnarray}
\end{linenomath*}
The first reduction of model complexity is via the Ott-Antonsen approach 
\cite{Ott08} for followers without Gaussian white noise but 
frequency heterogeneity $\gamma_0/\lambda_0$ with Cauchy-distributed 
frequency deviations $\nu_n$. In the thermodynamic limit $|P_\sigma|\to\infty$ {(keeping the ratios $w_{k\sigma}$ of the  partition sizes constant)} there exists a globally attractive invariant manifold on which the mean-fields $Z_\sigma$ evolve by a complex Riccati equation as
\begin{linenomath*}
\begin{equation}\label{Eq:OA_CommunityDyn}
\dot Z_\sigma = i\left(\bar{h}_\sigma Z_\sigma^2 
+ i\frac{\gamma_0}{\lambda_0} Z_\sigma + h_\sigma  \right) = F(Z_\sigma,h_\sigma).
\end{equation}
\end{linenomath*}
For large partition sizes,  Eqs. (\ref{Eq:Fields01}-\ref{Eq:OA_CommunityDyn})  provide a good 
description of the system dynamics, including an  
accurate description of the fluctuations of the mean-fields (see Supplementary Note 6). The effect 
of small noise $D_0/\lambda_0$ in the followers is comparable to the 
effect of frequency heterogeneity $\gamma_0/\lambda_0$. For small 
white noise, the Ott-Antonsen manifold is no longer invariant but one 
can derive a hierarchy of corrections to the dynamics 
(\ref{Eq:OA_CommunityDyn}) in increasing orders of the noise strength. 
To the zeroth order the effects of frequency heterogeneity  
and noise are identical \cite{Goldobin}. In fact, if the noise in the followers is white Cauchy noise, the equivalence of noise and frequency heterogeneity is exact \cite{tonjes2020low}.
\subsection*{Slow-fast dynamics}\label{Met:SFD}
If there is a large dynamical frequency gap $\Delta\Omega_k/\lambda_0\gg 1$  between the followers and an influencer, oscillators in the follower group experience an average force from the fast influencer. Conversely, if the followers are desynchronized, the mean-field of the followers vanishes and the influencer phases perform a drift diffusion process on the circle
\begin{linenomath*}
\begin{equation}\label{Eq:HubDriftDiff}
\dot\psi_k = \frac{\Delta\Omega_k}{\lambda_0} 
+ \sqrt{2D_k/\lambda_0}\tilde{\xi}_k(t)
\end{equation}
\end{linenomath*}
whereas the followers connected to only that influencer experience 
a stochastic forcing by the influencer phase
\begin{linenomath*}
\begin{equation}\label{Eq:LeafDrive}
\dot\vartheta_n = \frac{\gamma_0}{\lambda_0}\nu_n 
+ \sin\left(\psi_k-\vartheta_n-\alpha\right) + \sqrt{2D_0/\lambda_0}\tilde{\xi}_n(t).
\end{equation}
\end{linenomath*}
This forcing is multiplicative since $\sin\left(\psi_k-\vartheta_n-\alpha\right) 
= s_k \cos\vartheta_n - c_k \sin\vartheta_n$ with two uncorrelated but 
not independent random processes $s_k(t)=\sin(\psi_k-\alpha)$ and 
$c_k(t)=\cos(\psi_k-\alpha)$. The diffusion constants $D_s$ 
and $D_c$ for the integrated stochastic forces quantify an effective noise strength and can be calculated as the integral of the respective autocorrelation functions \cite{Risken}
\begin{linenomath*}
\begin{equation}\label{Eq:Deff}
\frac{D_s}{\lambda_0} = \frac{D_c}{\lambda_0} = 
\frac{\lambda_0}{2}\frac{D_k}{\Delta\Omega_k^2 + D_k^2} = 
\frac{\lambda_0}{2\Delta\Omega_k} \frac{q_k}{1+q_k^2}.
\end{equation} 
\end{linenomath*}
We present the details in Supplementary Note 5.  By changing the noise strength $D_k$, the effective noise strengths $D_s$ and $D_c$ have a maximum at  $D_k = \Delta\Omega_k$ or $q_k=D_k/\Delta\Omega_k=1$. At this noise value,  and for incoherent followers, the effect of noise-induced  synchronization \cite{Pimenova} is expected to be strongest. As  the amplitudes of $s_k$ and $c_k$ are bounded, when $D_k$ or the time scale separation $\Delta\Omega/\lambda_0$ are further increased the effective noise strengths go to zero. 

For $\Delta\Omega/\lambda_0\gg 1$, the system has  slow and  fast  dynamics and we can replace the influencer phases $z_k$ contributing to the force fields $h_\sigma$ (\ref{Eq:Fields02}) in each partition $\sigma$ by the expected values $G_k$ of $z_k$  subject to Langevin equation (\ref{Eq:HubDynamics}). {On the fast time scale of the influencers, the fields $h_k$ are changing very slowly and can assumed to be constant for the calculation of the $G_k$.} In this averaged dynamics, the influencers create  an average force $H_\sigma$ that follows the partition mean-fields  adiabatically. The slow dynamics of the partition mean-fields is thus given as
\begin{linenomath*}
\begin{eqnarray}
\dot Z_\sigma  &=& i\left(\bar{H}_\sigma Z_\sigma^2 
+ i\frac{\gamma_0}{\lambda_0} Z_\sigma + H_\sigma\right) = F(Z_\sigma,H_\sigma)
\label{Eq:HyperGraphDyn} \\
H_\sigma &=& \frac{e^{-i\alpha}}{2i} \frac{1}{|I_\sigma|}\sum_{k\in I_\sigma} 
G_k\left(\sum_{\sigma'} w_{k\sigma'} Z_{\sigma'};~\Lambda_k, q_k\right) .
\label{Eq:HyperGraphForce}
\end{eqnarray}
\end{linenomath*}
This corresponds to a hyper-graph with partitions $\sigma$ as  nodes and coupling functions $G_k$ for each edge $k$ of the  hyper-graph. General setups can be considered as well,  with intra and inter-partition coupling and connections between influencers. The absence of such connections  shows that the synchronization is indeed a noise-induced  effect.
\subsection*{Mean-field of the fast influencers}
The Langevin equation (\ref{Eq:HubDynamics}) for $z_k$ with 
constant fields $h_k$ is indeed a complex formulation of the noisy Adler equation \cite{Risken}
\begin{linenomath*}
\begin{equation}\label{Eq:HubPhaseLangevin}
\dot\psi = \frac{\Delta\Omega}{\lambda_0}\left(1 
+ 2\Lambda\textrm{Im}\left[h e^{-i\psi}\right]\right)  + \sqrt{2D/\lambda_0}\xi(t).
\end{equation}
\end{linenomath*}
The expected value $G$ of $z = \exp(i\psi)$ is  the first circular 
moment of the stationary distribution which has an expression 
as a continued fraction \cite{Risken} and evaluates to a ratio of 
confluent hypergeometric limit functions $_0F_1(o,x)$ \cite{Stratonovich}. 
It can be derived from the Fokker-Planck equation noting that the 
Fourier modes $p_k = \langle \exp(ik\psi)\rangle$ of the stationary 
distribution $p^{st}(\psi)$ are in a tridiagonal recurrence relation
\begin{linenomath*}
\begin{equation}\label{Eq:stat_Fourier_FPE}
0 = ik \Delta\Omega \left(\Lambda\bar{h}p_{k+1} + p_k 
+ \Lambda hp_{k-1} \right) - Dk^2p_k 
\end{equation}  
\end{linenomath*}
which is solved by a continued fraction. Defining $q=D/\Delta\Omega$ and
\begin{linenomath*}
\begin{equation}\label{Eq:Sox}
	s = \frac{|h|}{ih}, \quad o = 1 - i\frac{1}{q}, 
	\quad \textrm{and}\quad x = \frac{q}{\Lambda|h|} 
\end{equation}
\end{linenomath*}
we have $G=p_1$ and
\begin{linenomath*}
\begin{equation}\label{Eq:ContFrac}
	G = \frac{1}{o + \frac{x^{-2}}{(o+1) + \frac{x^{-2}}{(o+2) + \dots}}} 
	\frac{1}{sx} = \frac{_0F_1(o+1,x^{-2})}{_0F_1(o,x^{-2})}\frac{1}{sox}.
\end{equation}
\end{linenomath*}
\subsection*{Synchronization manifold and prediction of order parameter}
If all influencers have the same effective noise strength $q = q_k = \frac{D_k}{\Delta\Omega_k}$ and the same effective coupling strength $\Lambda=\Lambda_k=\frac{\beta\lambda_0}{\Delta\Omega_k}$, 
the synchronization manifold where  all partitions have identical 
mean-fields $Z_\sigma=Z=R^{i\Theta}$ is invariant under the 
averaged dynamics (\ref{Eq:HyperGraphDyn}) and (\ref{Eq:HyperGraphForce}) on the hyper-graph and we can write
\begin{linenomath*}
\begin{eqnarray}
	\dot Z  &=& i\left(\bar{H} Z^2 + i\frac{\gamma_0}{\lambda_0} Z 
	+ H\right) \label{Eq:SyncManifoldDyn} \\
	H &=& \frac{e^{-i\alpha}}{2i} G\left(Z;~\Lambda,q\right)	
	\label{Eq:SyncManifoldForce}
\end{eqnarray}
\end{linenomath*}
where $G(Z;~\Lambda,q)$ is defined as (\ref{Eq:ContFrac}) with
\begin{linenomath*}
\begin{equation}\label{Eq:SyncManifoldHubForce}
	h=\frac{e^{-i\alpha}}{2i}Z.
\end{equation}
\end{linenomath*}
In particular, because of rotational symmetry, the dynamics of the 
amplitude $R=|Z|$ does not depend on the angle $\Theta$ of the mean-field
\begin{linenomath*}
\begin{equation}\label{Eq:SyncManifoldAmplitudeDyn}
\dot R = \textrm{Re}\left[\frac{e^{-i\alpha}}{2}G(R;~\Lambda,q)\right]
\left(1-R^2\right) - \frac{\gamma_0}{\lambda_0} R.
\end{equation}
\end{linenomath*}
If the synchronization manifold is stable, the stable fixed points of this 
dynamics where $\dot R = 0$ approximate the average order parameter 
over all followers. From (\ref{Eq:SyncManifoldAmplitudeDyn}) we find that the level sets of the right-hand side of 
\begin{linenomath*}
\begin{equation}\label{Eq:SyncManifoldStationary}
\frac{\gamma_0}{\lambda_0} = \textrm{Re}\left[\frac{e^{-i\alpha}}{2}
G(R;~\Lambda,q)\right]\frac{1-R^2}{R}
\end{equation}
\end{linenomath*}
determine this average order parameter $R$ for any given $\gamma_0/\lambda_0$ 
implicitly. We show this prediction for $\gamma_0/\lambda_0=0.02$ and $\Lambda=0.51$ as a solid line in the lower right panel in Fig.\ref{Fig1}. Further resonance curves and maxima of $R$ for different heterogeneities $\gamma_0$ and different $\Lambda$ can be found in Supplementary Note 3.
\\\\\noindent
\textbf{Acknowledgements} We would like to thank C. Sagastiz\'abal, D. Eroglu, S. van Strien, D. Turaev, J. Lamb, and A. Pikovsky for enlightening discussions. This work was supported in parts by the DFG and FAPESP through the IRTG 1740/TRP 2015/50122-0, by the Center for Research in Mathematics Applied to Industry (FAPESP Cemeai grant 2013/07375-0) and grants 2015/04451-2, by the Royal Society London, CNPq grant 302836/2018-7, and by the Serrapilheira Institute (Grant No. Serra-1709-16124).
\\\\\noindent
\section*{Publication}

This work was published under Creative Commons CC BY license at
\\ \\
https://doi.org/10.1038/s41467-020-20441-4
\\ \\
together with Supplementary Notes, Movies and referee reports, and may be cited as
\\ \\
R. T\"onjes,C.E. Fiore and T. Pereira,{\em Coherence resonance in influencer networks}, Nat Commun {\bf 12}, 72 (2021)
\section*{Data Availability} 
The data that support the findings of this study are available from the corresponding author upon reasonable request.
\section*{Code Availability} Input files or sets of input parameters for Fortran 
as well as self-developed Python codes are available from the corresponding 
author upon request. Python code for full oscillator network simulations for an undirected scale free network and for the {\it C. elegans} directed neuronal network can be found at \\
https://github.com/rahleph/CR-on-Influencer-Networks.git
\section*{Author contributions} RT and TP wrote the text and developed the 
theory. RT made the slow fast approximation and numerical solutions of the 
special functions.  CEF  made simulations of Figure 1. TP and RT made the Figures. 
5
\section*{Competing interests.} Authors declare no competing interests. 
\bibliographystyle{naturemag}

\newpage
\section*{Supplementary Note 1: Example of transformation into canonical form}
We start with weakly non-identical ($\omega_0=2.0$, $\Delta\omega/\omega_0<10\%$), weakly coupled ($\lambda_0=0.01$, $\beta=10$) phase oscillators with moderately large shear ($c_0=0.5$) strong noise ($D=1.4$) in the hubs and weak noise (or weak frequency heterogeneity) in the followers ($D_0=2\times 10^{-4}$).
\begin{linenomath*}
\begin{eqnarray}
    \dot \vartheta_n &=& 2.0 + \frac{0.01}{\mu_n}\sum_{m} W_{nm}\left[ \sin(\vartheta_m-\vartheta_n-\alpha) + 0.5 \right] + \sqrt{4\times 10^{-4}}\xi_n  \\
    \dot \vartheta_{\rm hub} &=& 2.135 +  \frac{0.1}{\mu_{\rm hub}}\sum_{m} W_{{\rm hub}~m}\left[ \sin(\vartheta_m-\vartheta_{\rm hub}-\alpha) + 0.5 \right] + \sqrt{2\times 1.4} \xi_{\rm hub} \nonumber \\
\end{eqnarray}
\end{linenomath*}
After going into a co-rotating reference frame, where the follower mean natural frequency is zero $\vartheta\mapsto\vartheta - (\omega_0 +\lambda_0 c_0) t$ and a change of time scale $t\mapsto \lambda_0 t$, we obtain
\begin{linenomath*}
\begin{eqnarray}
    \dot\vartheta_n &=& \frac{1}{\mu_n}\sum_{m} W_{nm}\sin(\vartheta_m-\vartheta_n-\alpha) + \sqrt{2\times 0.02}\tilde{\zeta}_n  \\
    \dot \vartheta_{\rm hub} &=& 18\left[1 + \frac{10}{18}\frac{1}{\mu_{\rm hub}}\sum_{m} W_{{\rm hub}~m} \sin(\vartheta_m-\vartheta_{\rm hub}-\alpha)\right] + \sqrt{2\times \frac{140}{18} 18} \tilde{\zeta}_{\rm hub} \nonumber \\
\end{eqnarray} 
\end{linenomath*}
The effective dimensionless parameters for this system are $\Delta\Omega/\lambda_0 = 18$, 
$\Lambda=10/18$ and $q=140/18\approx 7$. The  large noise strength $D/\lambda_0=140$ is only large in units of $\lambda_0=0.01$. Here, $D=1.4$ is already much larger than the optimal noise strength $D_{\rm opt}=0.18$.
\section*{Supplementary Note 2: Hyper-graph}
\begin{figure}[t!]
\centering
\includegraphics[height=6cm]{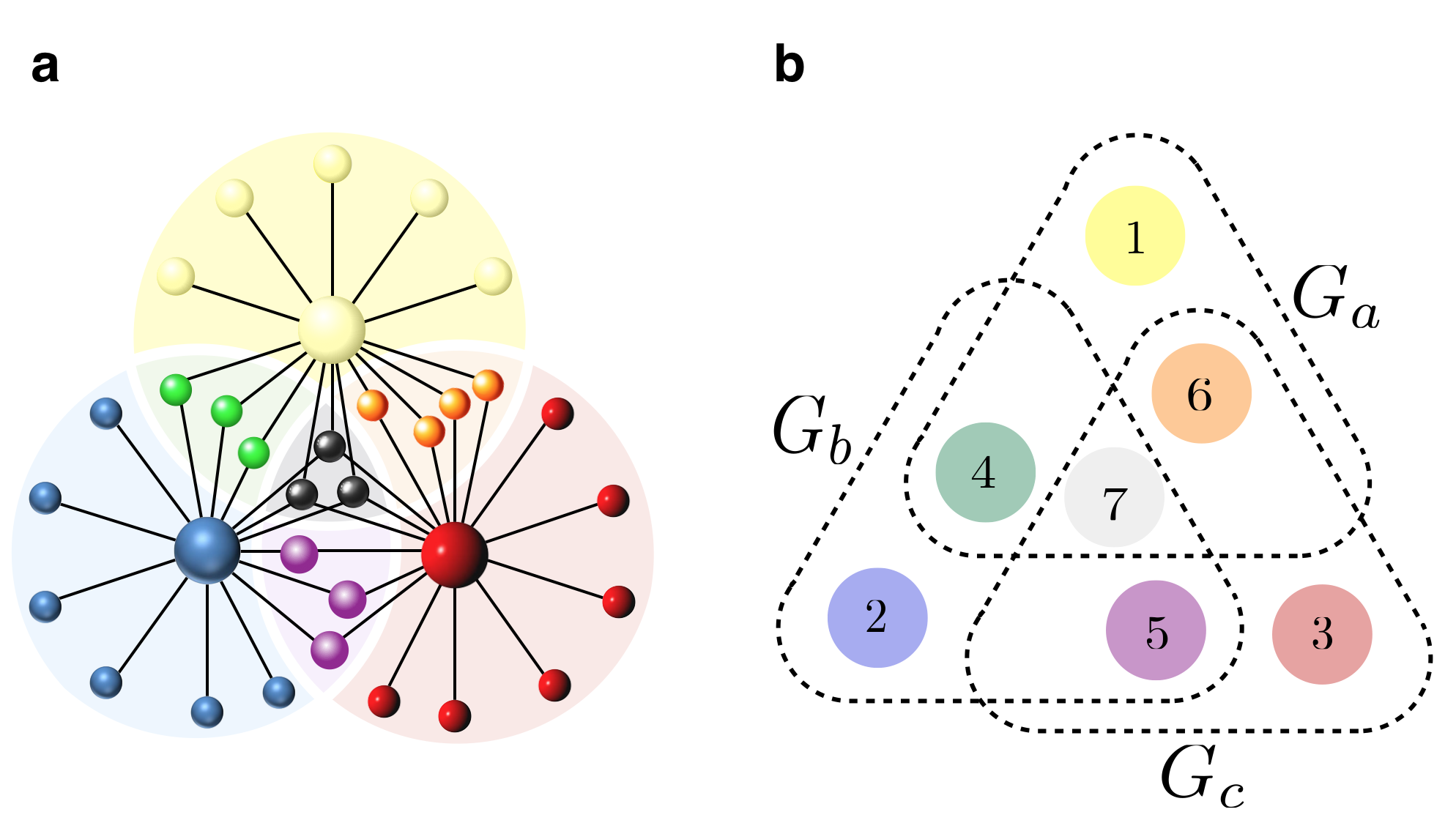}
\caption{{\bf Three-influencer network and hyper-network}. (a) network with 
three influencers. (b) a hyper-network of seven partitions connected by 
three  hyper-edges.}
\label{FigSupHyper7}
\end{figure}
We discuss the hyper-graph structure emerging at the macroscopic 
interaction of mean-fields. We will provide  illustration 
of an influencer network with seven partitions of equal size connected through three influencers. After the Ott-Antonsen reduction of the partition mean-field dynamics and 
after averaging over the fast influencers, the resulting mean-field equations 
have the following form
\begin{linenomath*}
\begin{eqnarray}
	\dot Z_1 &=& F\left(Z_1,G_a\right), \quad 	\dot Z_2 = F\left(Z_2,G_b\right), \quad
	\dot Z_3 = F\left(Z_3,G_c\right),  \\
	\dot Z_4 &=& F\left(Z_4,\frac{1}{2}G_a+\frac{1}{2}G_b\right),\quad \dot Z_5 = F\left(Z_5,\frac{1}{2}G_b+\frac{1}{2}G_c\right),\\
	\dot Z_6 &=& F\left(Z_6,\frac{1}{2}G_c+\frac{1}{2}G_a\right), \quad
	\dot Z_7 = F\left(Z_7,\frac{1}{3}G_a+\frac{1}{3}G_b+\frac{1}{3}G_c\right)
\end{eqnarray}
\end{linenomath*}
with coupling functions
\begin{linenomath*}
\begin{eqnarray}
	G_a &=& G\left(\frac{1}{4}Z_1 + \frac{1}{4}Z_4 + \frac{1}{4}Z_6 + \frac{1}{4} Z_7;\Lambda_a, q_a\right)  \\
	G_b &=& G\left(\frac{1}{4}Z_2 + \frac{1}{4}Z_4 + \frac{1}{4}Z_5 + \frac{1}{4}Z_7;\Lambda_b,q_b\right)  \\
	G_c &=& G\left(\frac{1}{4}Z_3 + \frac{1}{4}Z_5 + \frac{1}{4}Z_6 + \frac{1}{4}Z_7;\Lambda_c,q_c\right) 
\end{eqnarray}
\end{linenomath*}

\noindent
each connecting four partition mean-fields, and with Riccati dynamics
\begin{linenomath*}
\begin{equation}
	F(Z,G) = \left(\frac{e^{-i\alpha}}{2}G - \frac{\gamma_0}{\lambda_0} 
	Z -\frac{e^{i\alpha}}{2}\bar{G}Z^2 \right) 
\end{equation}
\end{linenomath*}
%
%
%
%
\section*{Supplementary Note 3: Impact of follower heterogeneity and noise}

In the main body of the manuscript we presented predictions and simulations with fixed values for frequency heterogeneity $\gamma_0/\lambda_0=0.02$ or noise strength $D_0/\lambda_0=0.02$ in the followers. The influence of these quantities on the coherence resonance can be seen in the mean-field analysis with slow-fast approximation. The frequency heterogeneity can be expressed as a function of all other system parameters, in particular the effective noise strength $q$ and the global order parameter $R$. The contour plot of $\gamma_0/\lambda_0$  as a function of $q$ and $R$ shows the resonance curves for constant follower heterogeneity  (see Supplementary Figure \ref{FigSup_Gamma}).
\begin{figure}[t!]
\centering
\includegraphics[height=6cm]{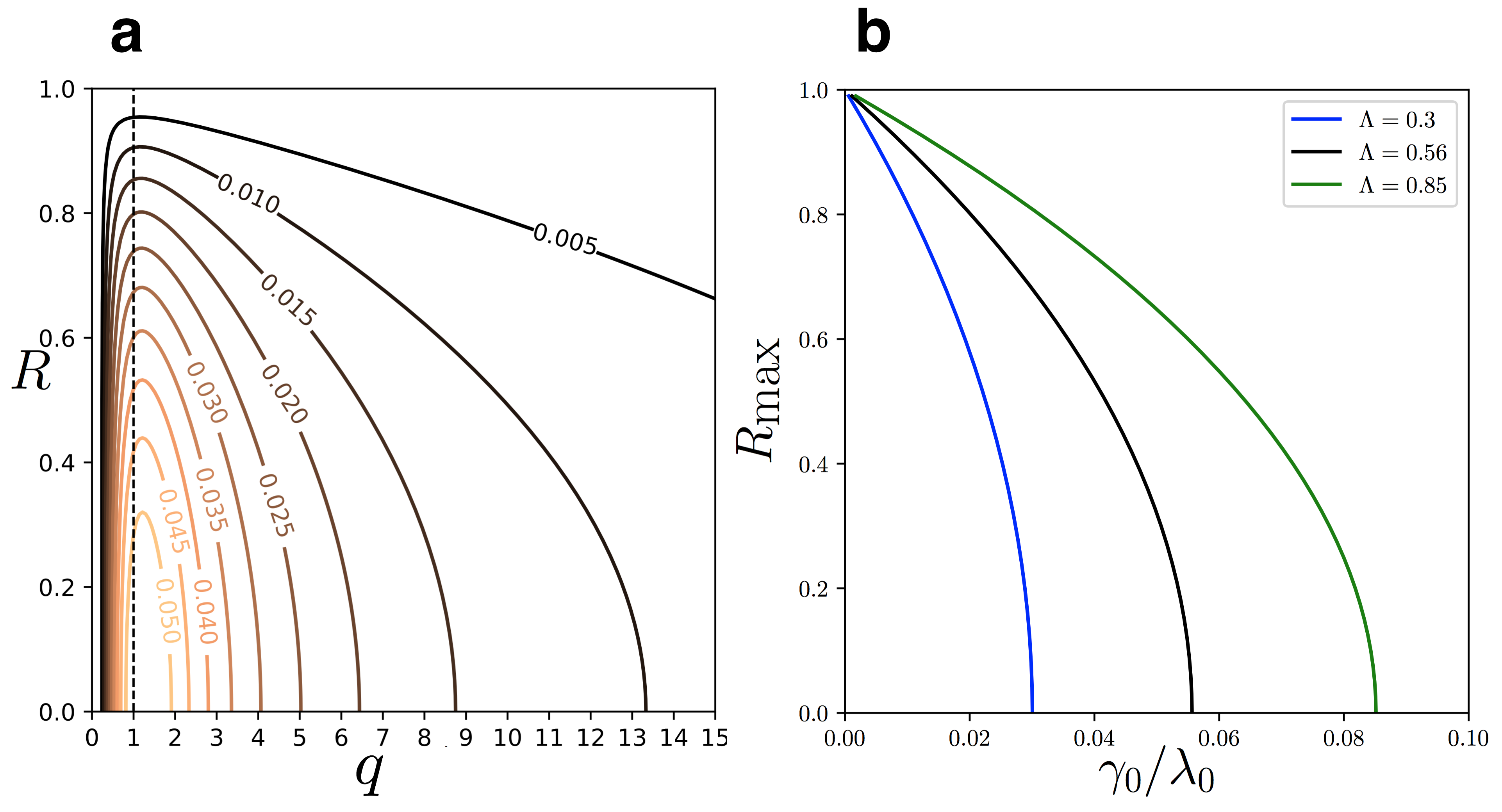}
\caption{
%
{\bf Effect of follower frequency heterogeneity on the coherence resonance}.
(a) shows resonance curves of the order parameter as a function of effective
noise $q$ in the influencers, with effective coupling $\Lambda=0.56$.
Shown are the level sets of the right-handside. of Eq. (26), which
correspond to different frequency
heterogeneities $\gamma_0/\lambda_0$ in the followers, as indicated by the
values in the labels. Using Gaussian white noise of equal strengths
$D_0/\lambda_0$ instead of Lorentzian frequency distribution for follower frequencies of width
$\gamma_0/\lambda_0$ results in similar resonance curves. (b) shows the maximum value of the order parameter as a function of the frequency heterogeneity for different values of effective coupling strength $\Lambda$.}
\label{FigSup_Gamma}
\end{figure}
%
\section*{Supplementary Note 4: Examples of coherence resonance}
We show three  examples of coherence resonance in Supplementary Figure \ref{FigSuppCR}.
For the three-influencer network motif, we choose each partition with
$100$ nodes and undirected connections. We show an influencer network of $3000$ nodes with $100$ influencers. Every node (also the influencers) is connected symmetrically to $k+1$ randomly selected influencers where $k$ is geometrically distributed with mean $10/6$. Since there are no direct links between followers, all connections have the weight $W_{mn}=1$. For the directed  hyperlink network of 1033 political weblogs after the 2004 US election, we choose the top 8 in-degree nodes as influencers. All nodes with zero out-degree have been removed. Phase coupling in the oscillator dynamics is realized in the opposite direction of the hyperlinks Ref. [11] of the main manuscript.   The time scale separation is $\Delta\Omega/\lambda_0=18$, $\beta=10$ ($\Lambda = 10/18$). The noise strength in the followers is $D_0/\lambda_0 = 0.02$ in the three-influencer network and $D_0/\lambda_0=0.04$ in the election blog network.

\begin{figure}[t!]
\centering
\includegraphics[width=12cm]{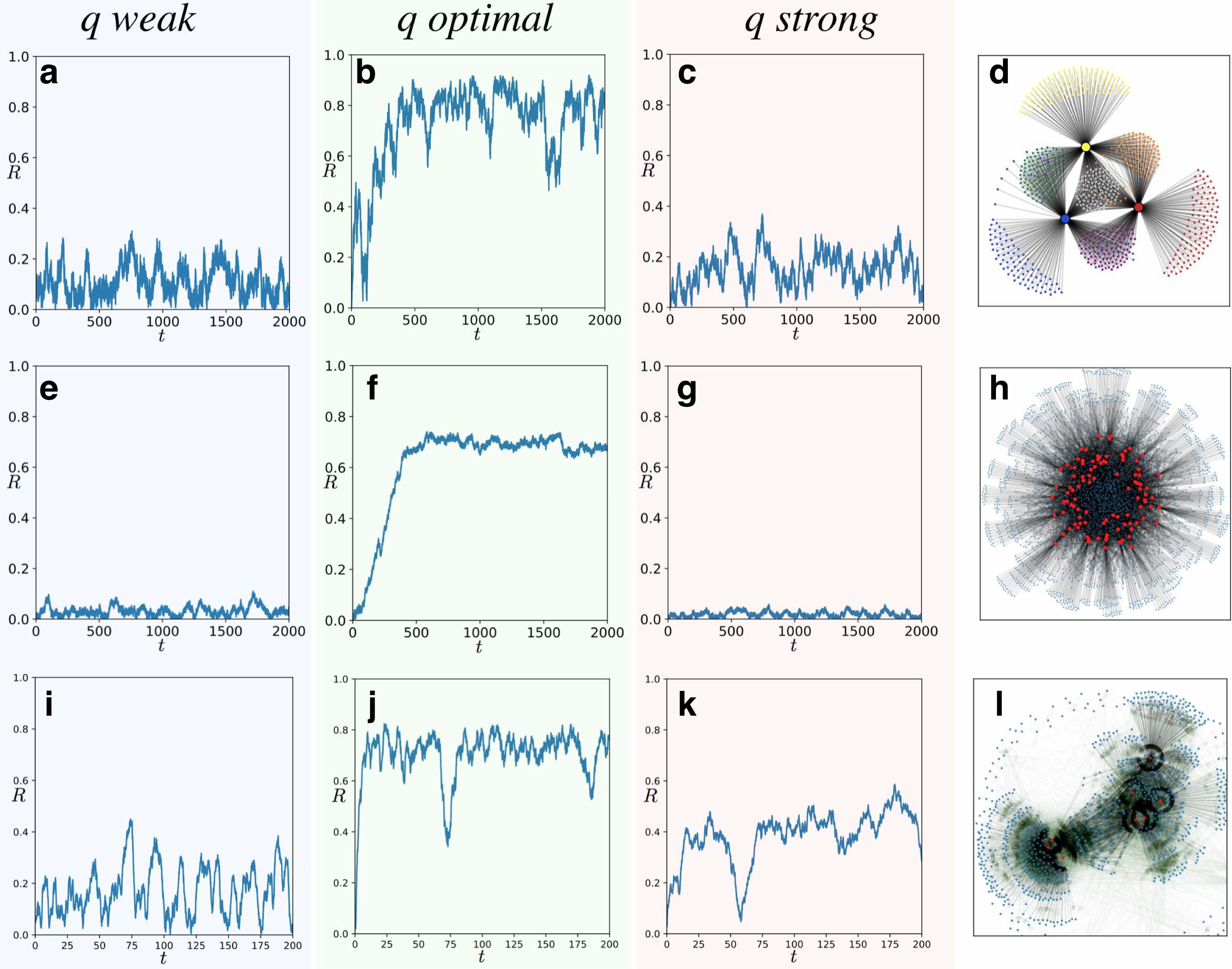}
\caption{{\bf Coherence resonance of the order parameter for different influencer networks}. 
We show the time series of the order parameter $R$ for three values of noise intensity 
in the influencers and the  network in the corresponding row. Parameters are $\Delta\Omega/\lambda_0=18$ and $\Lambda=10/18$.
In (a-d) we show the influencer network with three influencers and seven partitions as shown in SI Note 2, with $D_0/\lambda_0=0.02$. 
In (e-h) we show an influencer network with $3000$ nodes including $100$ influencers. 
In (i-l), we show the directed hyperlink network of $1033$ political weblogs after the 2004 US election (Ref. [11] of main manuscript) and the top 8-out degree nodes as influencers. Arrows indicate coupling direction which is in the opposite direction of the hyperlinks. The noise strength in the followers is $D_0/\lambda_0=0.04$.
The noise intensity in the influencers for the simulations in (a,e,i) is weak $q = 0.1$, in (b,f,j) optimal 
$q = 1.0$, and in (c,g,k) strong  $q = 10$.}
\label{FigSuppCR}
\end{figure}
\section*{Supplementary Note 5: Effective diffusion}
Here, we give an estimate of the effective noise strength for random forces $c(t)=\lambda_0\cos(\psi-\alpha)$ and $s(t)=\lambda_0\sin(\psi-\alpha)$ when $\psi$ is a drift-diffusion process
\begin{linenomath*}
\begin{equation}
\psi(t+\tau) = \psi(t) + \Omega\tau + W_D(\tau) \quad(\textrm{mod}~2\pi) 
\end{equation}
\end{linenomath*}
\noindent
on the circle with constant velocity $\Omega$, Brownian diffusion $W_D(\tau)$, and 
diffusion constant $D$. 
Such forces act on the follower phases coupled to influencers with phase $\psi$ when the global order parameter is zero, that is, in an incoherent state.

The conditional probability density $p(\psi,t+\tau|\psi_0,t)$ is a wrapped normal distribution 
$p_{WN}(\psi;\mu,\sigma^2)$ with mean $\mu=\psi_0+\Omega\tau$ and 
variance $\sigma^2 = 2D\tau$ , which in the limit $\tau\to\infty$ becomes 
a stationary, uniform distribution on the circle. The complex autocorrelation function
\begin{linenomath*}
\begin{equation}
C(\tau) = \lambda_0^2 \left\langle e^{-i\psi(t)}\cdot e^{i\psi(t+\tau)}\right\rangle_t  
\end{equation}
\end{linenomath*}
\noindent
is the sum of the autocorrelation functions of the two forces 
$c(t)=\lambda_0\cos(\psi-\alpha)$ and $s(t)=\lambda_0\sin(\psi-\alpha)$ 
in the real part and the cross-correlation function in the imaginary part. 
Replacing the time average by the average with respect to the conditional 
probability density and the average over the initial conditions $\psi_0$ 
with respect to the stationary probability density, the complex autocorrelation 
function has the form of the first circular moment of the wrapped normal 
distribution
\begin{linenomath*}
\begin{equation}
	C(\tau) = \lambda_0^2 \left\langle 
	e^{i\left(\Omega \tau + W_D(\tau)\right)} \right\rangle_{p_{_{WN}}} 
	= \lambda_0^2 e^{(i\Omega-D)\tau} .
\end{equation}
\end{linenomath*}

Because of the uniform stationary distribution, the autocorrelation 
functions of $c(t)$ and $s(t)$ are identical and both forces are 
uncorrelated; that is $\left\langle c(t)s(t) \right\rangle_t = 0$. The time 
integrals of the random forces are random variables that, according 
to the central limit theorem, have a variance that grows asymptotically 
linearly in time. The effective diffusion constant, which is half of the 
asymptotic speed of this growth, is the integral of the autocorrelation 
functions
\begin{linenomath*}
\begin{equation}
	D_{\rm eff} = \frac{\lambda_0^2}{2}\int_0^\infty \textrm{Re}
	\left[e^{(i\Omega-D)\tau}\right]d\tau  = \frac{\lambda_0^2}{2}
	\frac{D}{D^2+\Omega^2}
\end{equation}
\end{linenomath*}

Measuring time in units of $1/\lambda_0$, we obtain the
 expression in the Methods section of the manuscript.
\begin{figure}[t!]
\centering
\includegraphics[width=12cm]{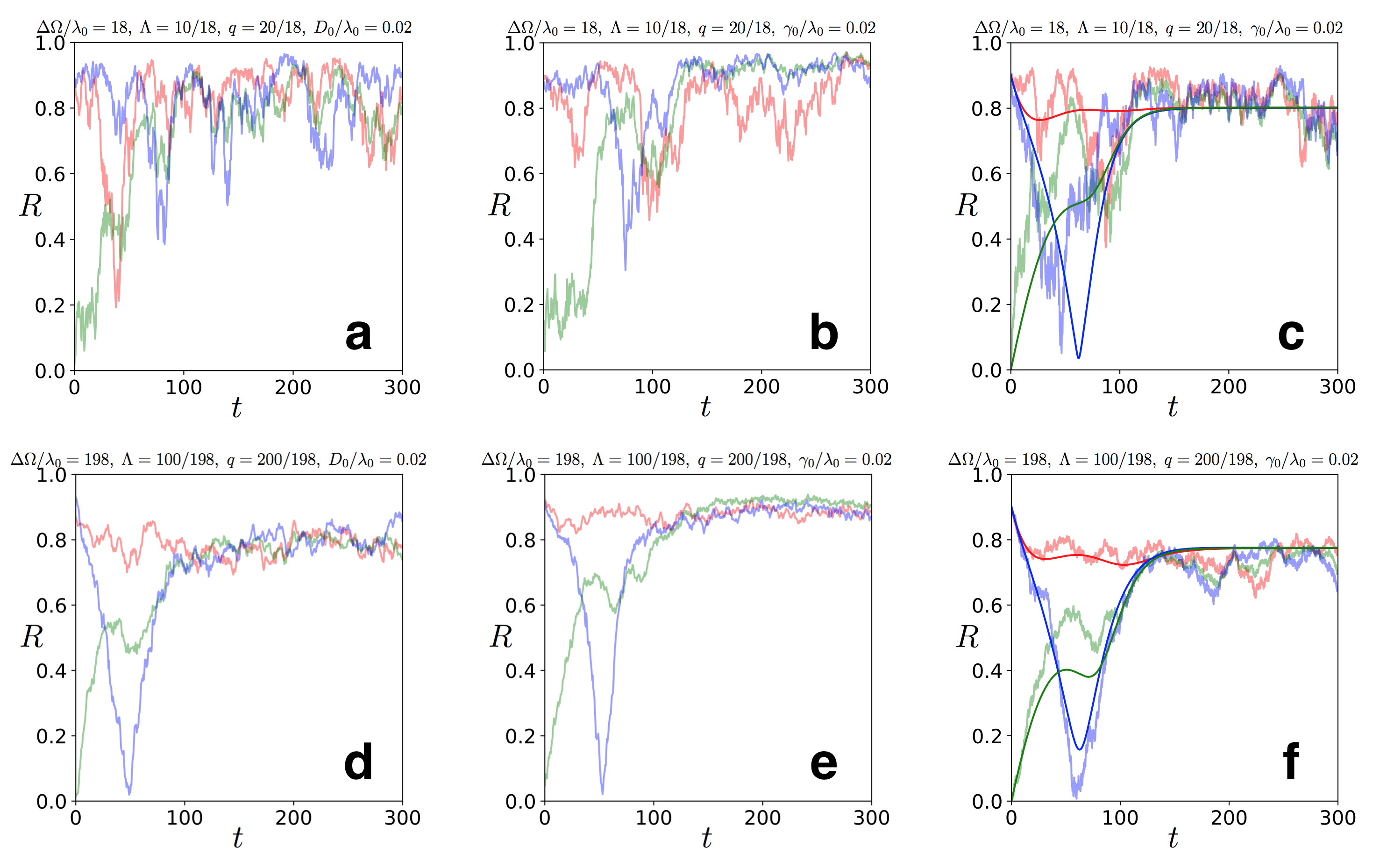}
\caption{{\bf Time series of the order parameters} $R_{\sigma}$ ($R_1$ red, $R_2$ green, $R_3$ blue) for the network in Figure 1 of the 
main manuscript with two influencers for (a-c) moderate time 
scale separation $\Delta\Omega/\lambda_0=18$, noise 
$q=20/18$ and reduced coupling strength $\Lambda=10/18$ 
and (d-f) large time scale separation $\Delta\Omega/\lambda_0=198$ 
with noise $q=200/198$ and $\Lambda=100/198$. Full 
network simulations with noise $D_0/\lambda_0=0.02$ in the 
followers are shown in (a) and (d). Full network simulations with 
frequency heterogeneity $\gamma_0/\lambda_0=0.02$ in the 
followers are shown in (b) and (e). In (c) and (f) we show 
simulations of the Ott-Antonsen mean-field equations under 
stochastic forcing by the influencers and of the averaged dynamics 
(smooth curves) obtained under the assumption of infinite time 
scale separation. The initial conditions are $Z_1 = 0.9\exp(i\pi/2)$, 
$Z_3 = 0.9\exp(-i\pi/4)$, $Z_3=0$, $\psi_a = \pi$ and $\psi_b=-\pi/2$. 
The drop in the order parameter $R_3$ is predicted by the averaged 
dynamics and can be observed in the full network simulations 
with noise $D_0$ or frequency heterogeneity $\gamma_0$.}
\label{FigSupTimeSeries}
\end{figure}

\section*{Supplementary Note 6: Comparison of reduced models}
The Ott-Antonsen ansatz for ensembles of phase oscillators 
with Lorenzian frequency distribution and under common forcing 
in the zeroth and in the first harmonics results is an exact expression for the dynamics of the mean-field. The forcing can be stochastic if interpreted as Stratonovich 
stochastic differential equation. In the thermodynamic limit 
$N\to\infty$ but for finite dynamical frequency gap 
$\Delta\Omega/\lambda_0$, the fluctuations in the influencer 
phases lead to fluctuations in the partition mean-fields that 
are comparable to the partition mean-field fluctuations in the finite size network dynamics. For a large dynamical frequency gap, the effective noise strength 
of the stochastic forcing vanishes. In the slow-fast approximation, 
the followers are only subject to the average forces 
from the influencers, which in turn depend adiabatically on the 
partition mean-fields. While fluctuations are completely absent 
in the averaged dynamics, the average forces depend continuously 
on the influencer effective noise strength $q=D/\Delta\Omega$. Indeed, while for $\Delta\Omega/\lambda_0 = 18$ the averaged dynamics describes the equilibrium order parameters near the optimal noise strength reasonably well, for much larger time scale 
separation $\Delta\Omega/\lambda_0 = 198$ even the transient 
to equilibrium is reproduced robustly Supplementary Figure \ref{FigSupTimeSeries}.

\end{document}